\documentclass[12pt,a4paper]{article}

\let\abs=\envert

\usepackage[polish]{babel}

\usepackage{amsfonts}
\usepackage{amsmath}
 
\begin{document}

\title{\bf Perfect Fluid Spacetimes With Two Symmetries}

\author{A. Szereszewski and J. Tafel}
\date{}
\maketitle

\noindent
Institute of Theoretical Physics, University of Warsaw,
Ho\.za 69, 00-681 Warsaw, Poland, email: tafel@fuw.edu.pl

\bigskip

\bigskip

\noindent
{\bf Abstract}.\\ 
\noindent
A method of solving perfect fluid Einstein equations with two commuting spacelike Killing 
vectors is presented. Given a spacelike 2-dimensional surface in the 3-dimensional nonphysical
Minkowski space the field equations reduce to a single nonlinear differential equation.  
An example is discussed.



\bigskip

\section{Introduction}
Spacetimes admitting two spacelike Killing vectors appear in various physical situations.
They can represent cylindrical waves, inhomogeneous cosmological models or gravitational 
 field around cosmic strings (see e.g. \cite{ver}  \cite{kr} \cite{akr} 
\cite{vil} and references therein).  
In late 70's Belinskii and Zakharov \cite{bz} described a soliton technique of solving the vacuum 
Einstein equations in the case with two spacelike Killing vectors.
Another technique allows to generate perfect fluid solutions from vacuum solutions 
\cite{wr}.
In this paper we propose a related  approach which doesn't require solving the vacuum equations. 
The method is based on the observation that some of the Einstein equations are identically satisfied if the
metric components are defined in a specific way by a 2-dimensional surface in 3-dimensional nonphysical Minkowski
space \cite{jt}. In this case  the field equations  reduce to a single differential equation. 
We demonstrate a usefulness of the method on an example.
 More examples  will be published elsewhere \cite{ts}.

\section{Reduction of the Einstein equations}  
Let $\cal M$ be a $4$ dimensional  manifold with metric tensor $g$ of the 
signature $+---$ and coordinates $x^\mu$ $(\mu=0,1,2,3$) . We are looking for solutions 
of the perfect fluid Einstein equations
 \begin{equation}
    R^\mu\,_\nu=\kappa[(\epsilon+p)u^\mu u_\nu-\frac{1}{2}(\epsilon-p)\delta^\mu\,_\nu]  \label{eineqn}
 \end{equation}
where $\kappa=8\pi G/c^4\,$, $p$ is the pressure, $\epsilon$ is the energy density and $u^\nu$ is the normalized
velocity  of the fluid, $u^\nu u_\nu=1$.
We assume that $g$ admits two commuting spacelike Killing vectors orthogonal to 2-dimensional surfaces. Then there exists a coordinate system such that  
 \begin{equation}
    g=g_{AB}dx^A dx^B+\rho\,n_{ab}dx^a dx^b   \label{metric}
 \end{equation}
where $A,B=0,1$ and $a,b=2,3$, functions $\rho$ , $g_{AB}$ , $n_{ab}$
depend only on the coordinates $x^0$ and $x^1$ , matrix $n=(n_{ab})$ is negative definite, $\rho>0$ and
\begin{equation}
   \det n=1 .   \label{detn=1} 
\end{equation}  
The Ricci tensor of $g$ can be obtained by a direct calculation or projection formalism of Geroch \cite{G} \cite{WR}. It takes the following form
 \begin{align}
  R_{AB}&=\frac{1}{2}R^{(2)}g_{AB}-\frac{1}{\rho}\rho_{\vert AB}+\frac{1}{2\rho^2}\rho_{,A}
          \rho_{,B}+\frac{1}{4}n_{ab,A}n^{ab}\,_{,B} \,, 
                                     \label{eqnRAB}\\
  R_{Aa}&=0 \, ,  \label{eqnRAa=0}\\
  R^a\,_b&=-\frac{1}{2\rho}(\rho_{,A}\delta^a\,_b+\rho\,n^{ac}n_{cb,A})^{\vert A}\ , \label{eqnRab} 
 \end{align} 
where $n^{-1}=(n^{ab})$ is the matrix inverse to $n$ , $R^{(2)}$ is the scalar curvature of the two 
dimensional metric $g_{AB}$ and $_{\vert A}$ denotes the covariant derivative with respect to the metric 
$g_{AB}$.

Under the assumption $\epsilon +p\neq 0$ the Einstein equations for $g$ yield $u^a=0$
and
 \begin{align}
  &\tilde{R}_{AB}=\kappa(\epsilon+p)u_A u_B \, ,    \label{eqn1}\\
  & \Box \rho=\kappa\rho (\epsilon-p) \,,   \label{treqn}\\ 
  &(\rho\, n^{-1}n\,^{,A})_{\vert A}=0 \,,  \label{eqnsur}
  \end{align}
where $\Box \rho=  \rho^{\vert A}\,_{\vert A}$ and
 \begin{equation}
     \tilde{R}_{AB}=R_{AB}+\frac{1}{2\rho}\Box\rho\,\, g_{AB} \, . 
 \end{equation}

Equations (\ref{eqn1}), (\ref{treqn}) can be solved by an appropriate choice of the quantities $\epsilon$, $p$ and $u^A$ provided 
 \begin{equation}  
   \det (\tilde{R}_{AB})=0   \label{detR=0}
 \end{equation}
and
\begin{equation}
\tilde{R}^A\,_A  \tilde R_{00}>0 \, .            \label{cond2}\\
 \end{equation}
Then the Einstein equations reduce to (\ref{eqnsur}), (\ref{detR=0}) and (\ref{cond2}).
This approach was successfully applied by several authors (see e.g. \cite{SV} \cite{MW}), usually in combination with the method of separation of variables. In this paper we propose to solve equation (\ref{eqnsur}) in terms of 2-dimensional surfaces and to treat equation (\ref{detR=0}) as an equation for a conformal factor in the metric $g_{AB}$. In this way the Einstein equations are reduced to (\ref{detR=0}) and (\ref{cond2}).

\section{Relation to surfaces}
Consider vector $\vec{n}$ with coefficients $n_i$ $(i=1,2,3)$ related to the matrix $n=(n_{ab})$ in the following 
manner
 \begin{equation}
           n  =\begin{pmatrix}
                  n_3 +n_1   &  n_2 \\
                     n_2     &  n_3-n_1 
               \end{pmatrix} \, .          \label{n_ab}
 \end{equation}
Since $n$ is negative definite and obeys condition (\ref{detn=1}) we have to assume that
 \begin{align}
   &n_3<0 \, ,                  \label{n3<0}\\
   &\vec{n}\,^2=-1 \, ,         \label{n2=1}
 \end{align}
where $\vec{n}\,^2=n_1\,^2+n_2\,^2-n_3\,^2$. In terms of $\vec{n}$ equation (\ref{eqnsur}) reads 
 \begin{equation}
   d(\rho\vec{n}\times \ast d\vec{n})=0 \, ,    \label{eqnH2}
 \end{equation} 
where $d$ is the exterior derivative and the star denotes the Hodge dual with respect to the metric $g_{AB}$.  
Equation (\ref{eqnH2}) is a variant of the Ernst equation  if $\Box \rho=0$ (then $p=\epsilon$). 
In contrast to (\ref{detR=0}) it is preserved, for any $\rho$, by the conformal transformations
of the metric $g_{AB}$
  \begin{equation}
     g_{AB}=\gamma e^{\psi}g'_{AB} \, ,        \label{conf}
  \end{equation}\\
where  $\gamma=\pm 1$ and $\psi$ is a function of  $x^0$ and $x^1$.  Because of this invariance we can first look for 
$\rho$, $\vec{n}$ and $g'_{AB}$ satisfying (\ref{eqnH2}) and then fit $\psi$ in order to fulfil 
(\ref{detR=0}) and (\ref{cond2}). 

Let $M^{2,1}$ be the 3-dimensional Minkowski space with coordinates $x^i$ and 
metric $\eta_{ij}$=diag(1,1,-1).  As shown in \cite{jt} every
solution of equation (\ref{eqnH2}) which satisfies condition
\begin{equation}
d\vec{n}\times d\vec{n}\neq 0\label{dn}
\end{equation}
 corresponds to a 2-dimensional spacelike 
surface $\Sigma$ in  $M^{2,1}$. Vector $\vec{n}$ can be identified 
with the normal vector of the surface.  Equation (\ref{eqnH2}) is  identically satisfied if $g_{AB}$ 
is proportional to the second fundamental form of $\Sigma$ and the function $\rho$ is related to 
the (necessarily negative) Gauss curvature $K$ of $\Sigma$ by
 \begin{equation}
   \rho=(-K)^{-1/2}\,. \label{K}
 \end{equation}

These observations lead to  the following theorem which describes a method of solving the Einstein equations with perfect fluid.\\

\bigskip
\noindent
{\bf Theorem.}\\
{\it Let $\Sigma$ be a spacelike surface in the 3-dimensional (nonphysical) Minkowski space with the negative
Gauss curvature $K$. Assume that 4-dimensional spacetime metric $g$ is given by (\ref{metric}), (\ref{n_ab}) and 
(\ref{conf}), where $\vec{n}$ is the unit normal vector of $\Sigma$ satisfying (\ref{n3<0}), the function $\rho$ 
is defined by (\ref{K}) and $g'_{AB}$ is proportional to the second fundamental form of $\Sigma$.   
Then the Einstein equations (\ref{eineqn}) reduce to equation (\ref{detR=0}) for $\psi$ and inequality (\ref{cond2})}. 

\bigskip 
 
In this method the pressure  and the energy  of the fluid are given in terms of the metric components as follows
 \begin{align}
  \kappa\, p&=\frac{1}{2}R^A\,_A \, ,                                         \label{p}\\
  \kappa\,\epsilon&=\frac{1}{2}R^A\,_A+\frac{1}{\rho}\Box \rho \, .  \label{e}
 \end{align}  
In most cases $p$ and $\epsilon$ do not undergo a barotropic equation of state (see \cite{ts} for counter examples of cosmological type). They satisfy  the dominant energy condition $\epsilon\geq \abs{p}$ if
 \begin{equation}
   \gamma \Box' \rho \geq 0 \    \label{del_rho}
 \end{equation}
($\Box'$ correspond to the metric $g'_{AB}$)
and inequality (\ref{cond2}) is replaced by 
 \begin{align}
&\tilde{R}^A\,_A>0  \label{cond3}\\
&\tilde R_{00}>0 \, .            \label{cond4}
 \end{align}
Condition (\ref{del_rho}) can be used
to fix $\gamma$.   
Conditions (\ref{cond3}), (\ref{cond4}) depend nontrivially on the conformal factor 
$\psi$. They  can be satisfied locally by an appropriate choice of the initial 
data for $\psi$ (note that equation (\ref{detR=0}) is of the second order in $\psi$, hence the initial data 
consists of two functions of one variable).

\section{Example}

As an example of application of Theorem we consider a rotational surface $\Sigma$ given  by 
\begin{equation}
x^i(\tau, x)=\big((\tau-\tau_0) \cos 2x,(\tau-\tau_0) \sin 2x, \tau^{-1}\big),
 \label{e1}
\end{equation}
where $\tau_0$ is a constant and $x^A=\tau ,x$ are the internal coordinates on $\Sigma$. In order to assure the spacelike character of $\Sigma$ we assume that 
\begin{equation}
\tau>\tau_0\geq 1\ .\label{e2}
\end{equation}
In this case the unit normal vector satisfying (\ref{n3<0}) reads
\begin{equation}
 \vec{n}=-\frac{1}{\sqrt{\tau^4-1}}\big(\cos 2x, \sin 2x, \tau^2\big)\label{e3}
\end{equation}
and 
the fundamental forms of $\Sigma$ are given by
\begin{equation}
 g_I=dx_id x^i=(1-\tau^{-4})d\tau^2+4(\tau-\tau_0)^2 dx^2\ , \label{e4}
\end{equation}
\begin{equation}
 g_{II}=d n_i d x^i=\frac{4(\tau -\tau_0)}{\sqrt{\tau^4-1}}\Big(\frac{d\tau^2}{2\tau (\tau-\tau_0)}-dx^2\Big)\ .\label{e5}
\end{equation} 
It follows from (\ref{K}) and the well known formula $K=\det{g_I}/\det{g_{II}}$ that 
\begin{equation}
 \rho=(\tau^4-1)\sqrt{\frac{\tau-\tau_0}{2\tau^3}}\ .\label{e6}
\end{equation}
According to Theorem one can relate with $\Sigma$ the following 4-dimensional metric (we choose $\gamma=1$)
 \begin{equation}
  \begin{split}
     g=&\rho\Big[ e^{\psi}\Big(\frac{d\tau^2}{2\tau (\tau-\tau_0)}-dx^2\Big)-\sqrt{\frac{\tau^2-1}{\tau^2+1}}\big(\cos x dy-\sin x dz\big)^2+\\
&-\sqrt{\frac{\tau^2+1}
                {\tau^2-1}}\big(\sin x dy+\cos x dz\big)^2\Big] \ . 
  \end{split}\label{e8}
 \end{equation}
For $\psi=\psi (\tau)$ the metric is a cosmological solution of Bianchi type $VII_0$ with the velocity field orthogonal to the hypersurfaces $\tau=$const. In this case conditions (\ref{detR=0}), (\ref{cond2}) reduce to $\tilde R_{11}=0$ and $\tilde R_{00}\neq 0$ . A particular solution of them is  given by
\begin{equation}
\psi=\frac{1}{2\tau_0}\log\frac{\tau-1}{\tau+1}-\frac{1}{\tau_0}\arctan \tau \ .\label{e7}
\end{equation}
Up to knowledge of the authors the metric defined by (\ref{e6})-(\ref{e7}) is a new non-tilted cosmological solution.
The corresponding energy  and pressure of the fluid can be 
obtained from formulas (\ref{p}) and (\ref{e}). Since the resulting expressions are quite complicated we describe here only their general properties when $\tau_0=2$. 
In this case metric (\ref{e8}) is regular for all $\tau>0$. For large values of $\tau$ the cosmic time $t$ is proportional to $\tau^{3/2}$ and $g$ tends to the flat Robertson-Walker metric. The energy density is a monotonically decreasing function of $\tau$ such that 
\begin{equation}
   \epsilon \xrightarrow[\tau\rightarrow {\tau_0}]{} \infty \, , \qquad
\epsilon \xrightarrow[\tau\rightarrow\infty]{} 0\ .    \label{ep}
 \end{equation}
The ratio $p/\epsilon$ is bounded by 1 and -1 (thus, the energy dominant condition is satisfied). It reaches a minimal value lower than $-\frac{1}{3}$ and it  has the following asymptotes
 \begin{equation}
   \frac{p}{\epsilon} \xrightarrow[\tau\rightarrow {\tau_0}]{} 1 \, , \qquad
\frac{p}{\epsilon} \xrightarrow[\tau\rightarrow\infty]{} -\frac{1}{3}\ .    \label{p/e}
 \end{equation}
This type of behaviour of energy and pressure may be interesting from the point of view of the modern cosmology (see e.g. \cite{Z}). 

When $\psi_{,x}\neq 0$ the metric (\ref{e8}) can be interpreted as an inhomogeneous cosmological solution. In this case equation (\ref{detR=0}) cannot be easily solved.

\bigskip
\bigskip

\noindent {\bf Acknowledgments}.  This work was partially supported by
the Polish  Committee for Scientific Research (grant 2 P03B 127 24 and 2 P03B 036 23).

\end{document}